\documentclass{aa}  

\usepackage{graphicx}
\usepackage{txfonts}
\usepackage{amsmath}
\usepackage{amssymb}
\usepackage{natbib}

\graphicspath{{./figures/}}
\bibpunct{(}{)}{;}{a}{}{,}
\newlength{\minuslength}
\begin{document}

   \title{ALMA observations of doubly deuterated water: Inheritance of water from the prestellar environment}


   \author{S. S. Jensen\inst{1}\thanks{\email{sigurd.jensen@nbi.ku.dk}}
                \and J. K. J{\o}rgensen\inst{1}
          \and L. E. Kristensen\inst{1}
          \and A. Coutens\inst{2,3}
          \and E. F. van Dishoeck\inst{4,5}
          \and K. Furuya\inst{6} 
          \and D. Harsono\inst{4,7}
          \and M. V. Persson\inst{8}}

   \institute{Niels Bohr Institute \& Centre for Star and Planet Formation, University of Copenhagen, {\O}ster Voldgade 5-7, DK-1350 Copenhagen K, Denmark
	\and Laboratoire d'Astrophysique de Bordeaux, Univ. Bordeaux, CNRS, B18N, all{\'e}e Geoffroy Saint-Hilaire, 33615 Pessac, France
	\and Institut de Recherche en Astrophysique et Plan{\'e}tologie, Universit{\'e} de Toulouse, UPS-OMP, CNRS, CNES, 9 av. du Colonel Roche, 31028 Toulouse Cedex 4, France
	\and Leiden Observatory, Leiden University, PO Box 9513, 2300 RA Leiden, The Netherlands
	\and Max--Planck Institute f{\"u}r extraterrestrische Physik (MPE), Giessenbachstrasse, 85748 Garching, Germany
   \and National Astronomical Observatory of Japan, Osawa 2--21--1, Mitaka, Tokyo 181--8588, Japan
    \and Institute of Astronomy and Astrophysics, Academia Sinica, No. 1, Sec. 4, Roosevelt Road, Taipei 10617, Taiwan, R. O. C.
	\and Department of Space, Earth and Environment, Chalmers University of Technology, Onsala Space Observatory, 439 92 Onsala, Sweden}

   \date{Draft date: \today}

  \abstract
  {Establishing the origin of the water D/H ratio in the Solar System is central to our understanding of the chemical trail of water during the star and planet formation process. Recent modeling suggests that comparisons of the D$_2$O/HDO and HDO/H$_2$O ratios are a powerful way to trace the chemical evolution of water and, in particular, determine whether the D/H ratio is inherited from the molecular cloud or established locally.}
  {We seek to determine the D$_2$O column density and derive the D$_2$O/HDO ratios in the warm region toward the low-mass Class 0 sources B335 and L483. The results are compared with astrochemical models and previous observations to determine their implications for the chemical evolution of water.}
  {We present ALMA observations of the D$_2$O $1_{1,0}$--$1_{0,1}$ transition at 316.8~GHz toward B335 and L483 at $\lesssim 0.\!\!^{\prime\prime}5$ ($\lesssim 100~$au) resolution, probing the inner warm envelope gas. The column densities of D$_2$O, HDO, and H$_{2}^{18}$O are determined by synthetic spectrum modeling and direct Gaussian fitting, under the assumption of a single excitation temperature and similar spatial extent for the three water isotopologs.} 
  {D$_2$O is detected toward both sources in the inner warm envelope. The derived D$_2$O/HDO ratio is $(1.0\pm0.2)\times10^{-2}$ for L483 and $(1.4\pm0.1)\times10^{-2}$ for B335. These values indicate that the D$_2$O/HDO ratio is higher than the HDO/H$_2$O ratios by a factor of $\gtrsim2$ toward both sources.}
  {The high D$_2$O/HDO ratios are a strong indication of chemical inheritance of water from the prestellar phase down to the inner warm envelope. This implies that the local cloud conditions in the prestellar phase, such as temperatures and timescales, determine the water chemistry at later stages and could provide a source of chemical differentiation in young systems. In addition, the observed D$_2$O/H$_2$O ratios support an observed dichotomy in the deuterium fractionation of water toward isolated and clustered protostars, namely, a higher D/H ratio toward isolated sources.} 


   \maketitle
%
\section{Introduction}
Addressing when and how water is synthesized during the formation of stars and planetary systems is essential to our understanding of the delivery of water, and other volatiles, to young planets. Water is a requirement for life as we know it and therefore central for the habitability of planets \citep[e.g.,][]{chyba2005}. Additionally, water acts as a coolant in molecular gas and is a major constituent of the solid mass reservoirs in protoplanetary disks \citep[e.g.,][]{dishoeck2014, karska2018}.

\begin{figure*}[!htb]
\centering
\resizebox{0.75\hsize}{!}
        {\includegraphics{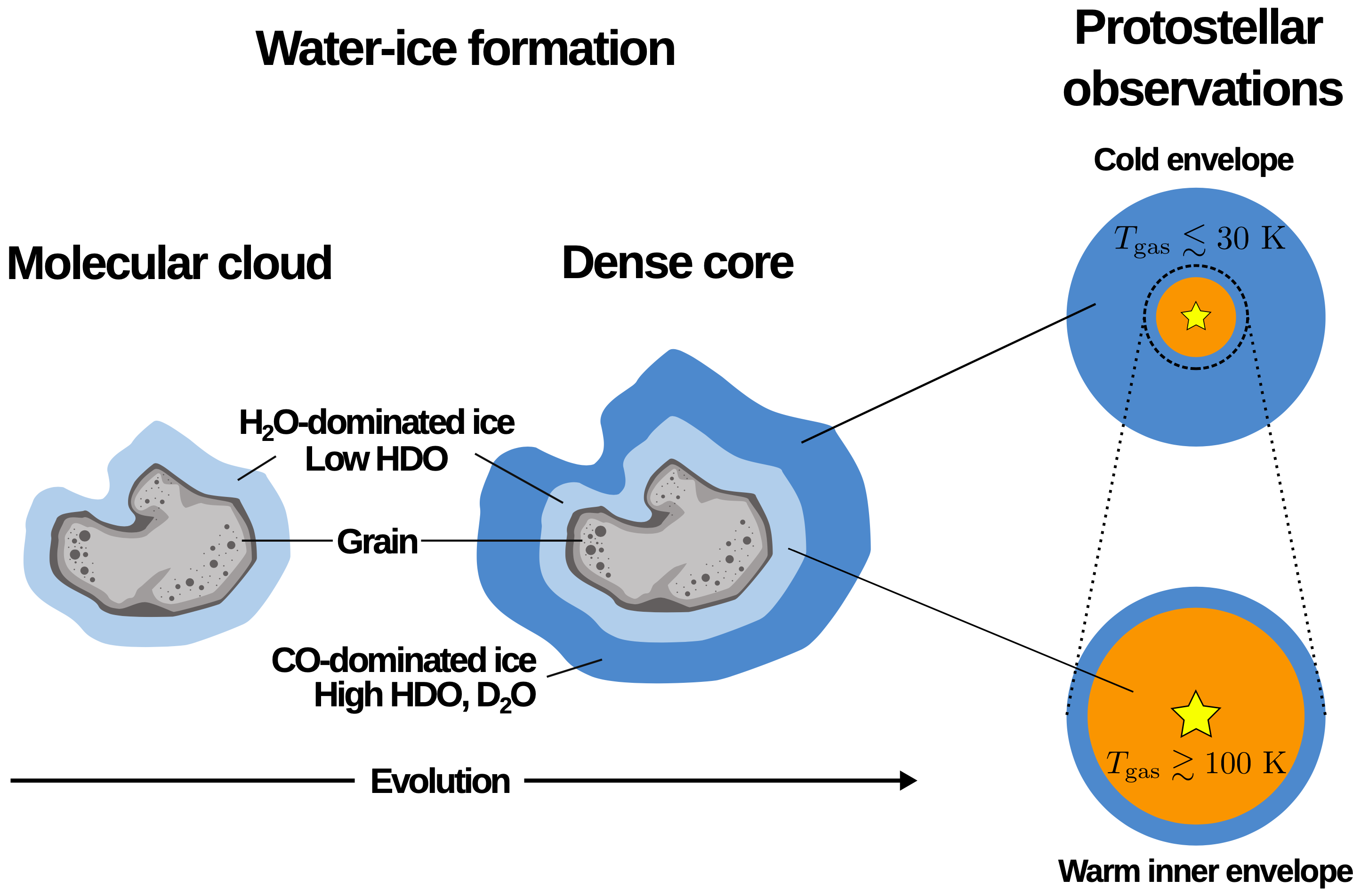}}
  \caption{Schematics of the predicted ice structure in the prestellar core stage (left) and the observed water deuterium fractionation toward embedded low-mass protostars (right). The outer layer of the ice mantle, which has high water D/H ratio, may stem from nonthermal desorption in the cold envelope or cold gas-phase formation. Meanwhile, the whole mantle is sublimated in the hot corino, lowering the HDO/H$_2$O ratio. The schematical structure is proposed by \cite{furuya2016}. The figure is adopted from \citet{vandishoeck2020}.
  .} 
     \label{fig:schematic}
\end{figure*}

Water is among the most abundant molecules in the interstellar medium and has been observed at all major stages of star and planet formation. In prestellar cores, water is a major constituent of the thick ice caps with $X_\mathrm{ice}\sim10^{-4}$ with respect to H$_2$ \citep[e.g.,][]{whittet1996, bergin2002, boogert2015}. In the inner warm envelope toward Class 0 sources, water ice is thermally desorbed, and the derived gas-phase water abundance reaches $X_\mathrm{gas}\sim10^{-7}$--$10^{-4}$ \citep[e.g.,][]{ceccarelli2000, persson2016}. In outflows, water is sputtered off the grains or synthesized directly in the gas phase, and the abundances can reach $X_\mathrm{gas}\sim10^{-6}$--$10^{-4}$ \citep[e.g.,][]{kristensen2017}. Meanwhile, a recent search for water toward nearby Class I sources shows that gas-phase water abundances are low in young disks($X_\mathrm{gas}\lesssim10^{-6}$), indicating that most of the water remains locked in ice at this stage \citep{harsono2020}. These observations provide vital constraints on the chemical evolution of water and reveal that water is formed as ice at high abundance in molecular clouds and prestellar cores \citep{vandishoeck2020}. This is confirmed by models of the chemical evolution during star formation \citep[e.g.,][]{taquet2013model, furuya2016, coutens2020}. However, a central question remains as to what extent is water in protoplanetary disk inherited from the prestellar cloud or synthesized locally in the disk \citep[e.g.,][]{visser2009, cleeves2014}. If prestellar inheritance is the dominant source of water in the protoplanetary disk, then the water chemistry is set by the conditions in the local cloud environment (e.g., the temperature, radiation field, and duration). This could promote some chemical differentiation in planetary systems. Conversely, if local processing in the disk is significant, then local conditions in the disk may strongly influence the chemistry of water and other volatiles.

By studying the deuterium fractionation of water, we can determine how the physical conditions affect the water chemistry and track the chemical evolution during star and planet formation \citep[e.g.,][]{ceccarelli2014}.
Deuterium fractionation occurs as a consequence of several chemical processes that depend sensitively on the temperature, density, visual extinction, and ionization sources \citep{caselli2012}. 
At low temperatures ($\lesssim50~$K), the prominent pathway is the gas-phase exchange reaction H$_{3}^{+}$ + HD $\rightleftharpoons$ H$_2$D$^{+}$ + H$_2$ + $\Delta E$, where $\Delta E $ depends on the spin state of the involved reactants. This forward reaction is endothermic, and the backward reaction is inhibited at low temperatures, thereby leading to an enrichment of H$_2$D$^{+}$ relative to H$_{3}^{+}$. H$_2$D$^+$ (and D$_2$H$^+$, D$_{3}^{+}$) dissociatively recombines with free electrons to form atomic D, thus increasing the local atomic D/H ratio in the gas phase and ultimately on dust grain surfaces where water and other molecules are formed through hydrogenation \citep{tielens1983, roberts2003}. Detections of gas phase H$_2$D$^{+}$ confirm that this enrichment is occurring in the gas phase in prestellar cores \citep{caselli2003, vastel2004}.

Observations of HDO and H$_{2}^{18}$O toward Class 0 source show that the degree of deuterium fractionation varies on different spatial scales \citep[e.g.,][]{persson2014}. Single-dish observations, some of which may arise in the cold extended envelope, reveal high D/H ratios of water; HDO/H$_2$O ratios range from 10$^{-3}$ to 10$^{-2}$ \citep{stark2004, parise2005, coutens2012}. 
Meanwhile, interferometric observations targeting the inner warm region have found much lower water deuteration; in these observations HDO/H$_2$O ratios range from $10^{-4}-10^{-3}$ \citep{jorgensen2010a, taquet2013observation, persson2014, jensen2019}. The first detection of interstellar D$_2$O was presented by \cite{butner2007}, who derived a D$_2$O/HDO ratio of 1.7$\times10^{-3}$ toward IRAS16293--2422 based on single-dish observations. In a later work, \cite{coutens2012} found a higher D$_2$O/HDO ratio of $\sim$0.06 toward the same source using \emph{Herschel} HIFI data. Interferometric detections of D$_2$O on inner envelope scales prior to this work are limited to \cite{coutens2014}, who found a high D$_2$O/HDO ratio of $\sim 1\%$ and D$_2$O/H$_2$O $\sim$ 2$\times$10$^{-5}$ toward NGC1333 IRAS~2A. The current observational understanding is outlined in Fig. \ref{fig:schematic}.


The observed difference in D/H ratios on different spatial scales can be understood as a result of the chronological evolution of water chemistry during star formation \citep{dartois2003, taquet2013model, furuya2016}. Before cloud collapse, water ice is formed in the molecular cloud with a lower deuterium fraction, as a result of higher temperatures and moderate extinction in this phase ($T\sim20$~K, $A_\mathrm{v}\lesssim5$~mag). Later, as the cloud core cools and contracts, the extinction increases and CO is depleted \citep[e.g.,][]{caselli1999}. During this phase, the deuterium fractionation becomes highly efficient, and a second generation of highly deuterated water ice is formed. 
This chronology leads to two distinct components in the interstellar ice: the bulk ice mantle consists of a water-rich ice with lower D/H ratio, and the surface consists of a mix of highly deuterated water, CO, and complex organics (see Fig. \ref{fig:schematic}). This scenario can explain the observed deuterium fractionations: on larger spatial scales, the bulk water reservoir remains in the solid phase, and the observed gas phase primarily consists of water with a high D/H ratio. This gas-phase water is either photo-desorbed from the ice surface layers or formed directly in the gas phase at low temperature and high extinction, that is, with a high D/H ratio \citep{taquet2013model, furuya2016}. Meanwhile, in the inner warm gas, where the ice is entirely sublimated, the gas-phase HDO/H$_2$O ratio of water is lowered, since the ice reservoir with a lower D/H ratio is released from the dust grains.
The similarity between D$_2$O/HDO ratios on larger and smaller scales support this scenario; the abundance of deuterated water isotopologs is low in the bulk water ice, and consequently the ratio is largely unaffected by the sublimation of the bulk water ice. 

A consequence of the above scenario for water formation is that the D/H ratio of water could depend on the physical conditions in the molecular cloud and prestellar core prior to and during the protostellar collapse. Lower temperatures or a longer duration of the deeply embedded phase would increase the D/H ratio observed in the inner warm region. Conversely, a faster collapse or warmer gas temperature could lower the D/H ratio in the inner warm region. This would imply a relation between the bulk water D/H ratio and the dynamics of the formation environment. 
Isolated protostars may collapse more slowly than clustered counterparts because the stability of the cloud is determined solely by the internal cloud structure \citep{ward-thompson2007, enoch2008}. Moreover, the temperature in clustered regions could be higher owing to, for instance, more external radiation or shock heating \citep{krumholz2014, hocuk2017}. Either of these scenarios would enhance the D/H ratio toward isolated protostars in contrast to clustered protostars, in which external factors can influence the physical conditions. Such a differentiation was recently suggested based on ALMA observations of HDO and H$_{2}^{18}$O in the warm inner envelope toward three isolated protostars \citep{jensen2019}. 
If the enhanced HDO/H$_2$O ratio is a consequence of environmental effects, then the same might also apply to D$_2$O.


In this work, we present the first ALMA detections of D$_2$O in the inner warm region toward the isolated low-mass Class 0 protostars B335 and L483. We combine the D$_2$O detections with previous detections of HDO and H$_{2}^{18}$O toward the sources to derive D$_2$O/HDO, D$_2$O/H$_2$O and HDO/H$_2$O ratios consistently. This work aims to constrain the chemical evolution of water between the molecular cloud and the inner warm envelope toward Class 0 sources. Furthermore, accurate determination of the D/H ratio in the warm envelope allow for comparison between the deuterium fractionation in protoplanetary disk and warm envelopes toward Class 0 sources, thereby establishing a possible link between these regimes.

The paper is organized as follows. In Section 2, the observations and data reduction are presented. The data analysis is presented in Section 3, and the results are discussed in Section 4. Throughout the paper we refer to NGC1333 IRAS~2A as IRAS~2A. We denote HDO/H$_2$O as f$_\mathrm{D1}$ and D$_2$O/HDO as f$_\mathrm{D2}$, and use both terms interchangeably. Furthermore, we refer to the inner warm gas component, where $r \lesssim 100$~au and $T \gtrsim 100$~K, as the warm envelope.

\section{Observations} \label{sec:2}

\begin{table*}
\centering\caption{Observation log.}             
\label{table:observations}
\centering          
\begin{tabular}{c c c c c c c}  
\hline \hline       
            \noalign{\smallskip}
Source &   Date & Phase calibrator & Bandpass calibrator & Max. baseline~(m) & n$_{\mathrm{antennas}}$ & $v_\mathrm{LSR} $(km/s)\\  
            \noalign{\smallskip}
\hline                   
            \noalign{\smallskip} 
            L483  & 2019 October 29 &  J1743$-$0350 & J1924$-$2914  & 697 & 42 & 4.5 \\
            
            B335 & 2019 October 8 & J1938$+$0448 & J1924$-$2914 & 784 & 42 & 8.3  \\

            \noalign{\smallskip} 
\hline                                    
\end{tabular}
\end{table*}

\begin{figure}
  \centering
  \includegraphics[width=\hsize]{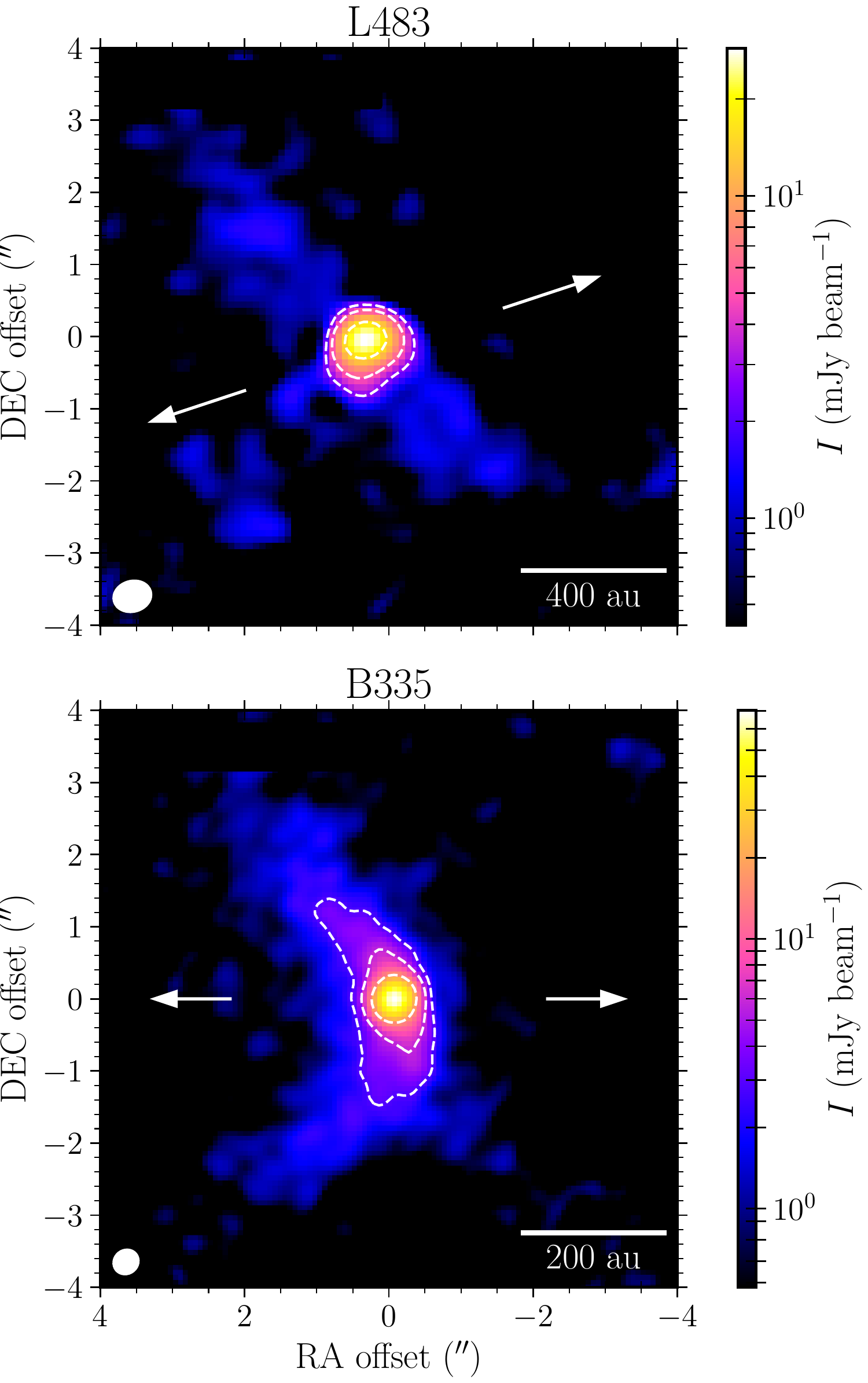}
      \caption{Continuum emission at 301.7~GHz toward L483 and B335. The map shows emission above 1$\sigma_\mathrm{rms}$ with white contours at 5$\sigma_{\mathrm{rms}}$, 10$\sigma_{\mathrm{rms}}$, and 30$\sigma_{\mathrm{rms}}$. Linear scales are indicated in the lower right corner, assuming a distance of 100~pc to B335 and 200~pc to L483 \citep[see][and references therein]{jensen2019}. The gray arrows indicate the direction of outflows that may be perturbing the dust distribution. The synthesized beam size is shown in the lower left corner.}
         \label{fig:cont}
  \end{figure}

\begin{figure*}
\centering
\resizebox{0.9\hsize}{!}
        {\includegraphics{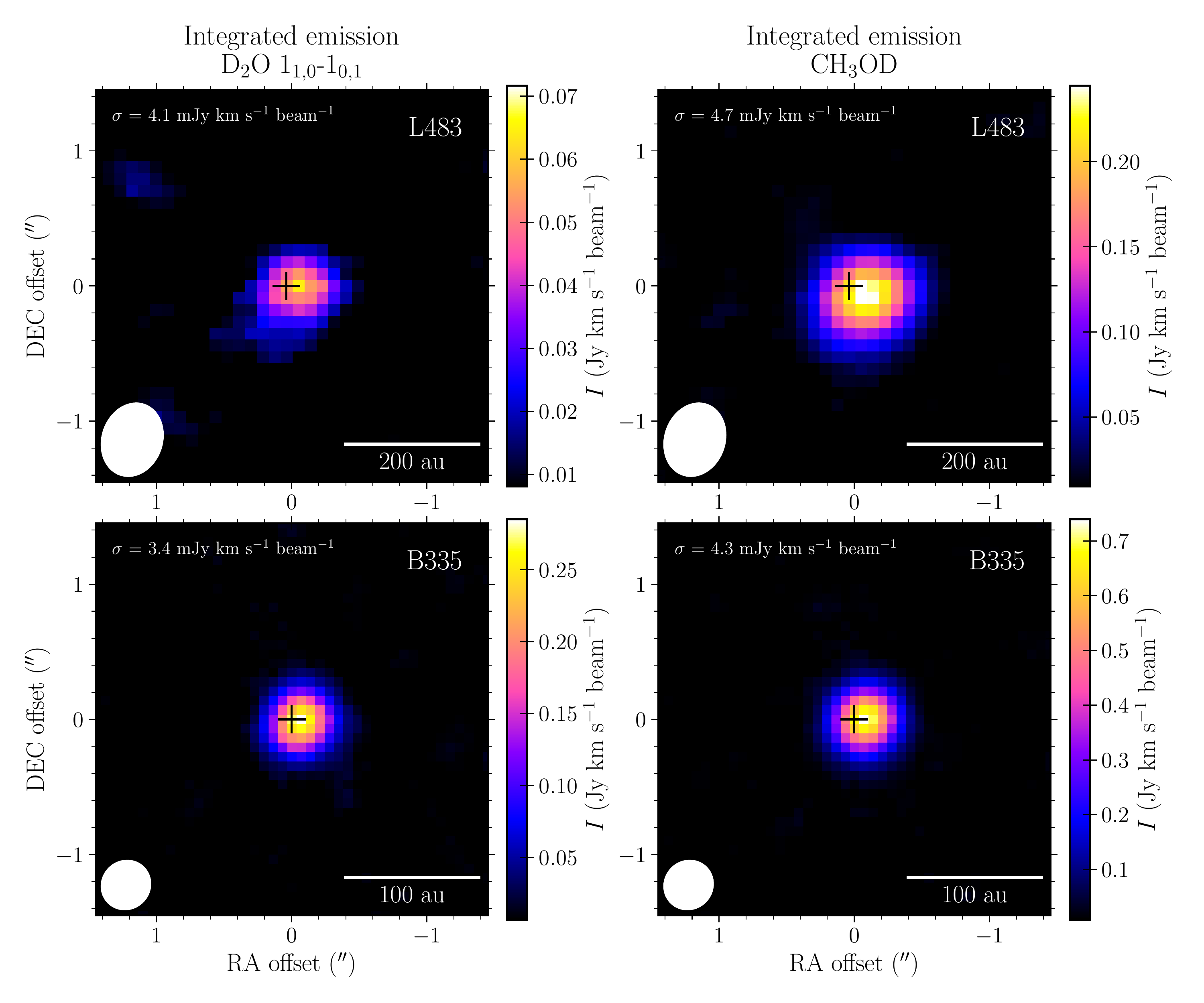}}
  \caption{Moment 0 maps of the D$_2$O and CH$_3$OD emission, excluding channels where the emission is blended. Maps show emission above $3\sigma$, where $\sigma = \sigma_{\mathrm{rms}} \times N_\mathrm{channels}^{0.5} \times d\varv$; $d\varv$ is the channel width and $N$ is the number of collapsed channels. The black cross indicates the continuum peak position.}
     \label{fig:moment0}
\end{figure*}

The low-mass protostars B335 and L483 were observed with ALMA during Cycle 7 (PI: Sigurd S. Jensen, projectid: 2019.1.00720.S). Information on the observation dates and calibration sources can be found in Table \ref{table:observations}. In the present analysis, we also utilize additional spectral windows from the data presented in \cite{jensen2019} toward both sources; this latter work gives a detailed description of those observations. 

The spectral setup targeted the D$_2$O $1_{1,0}$--$1_{0,1}$ transition at 316.8119~GHz in the ALMA band 7.
The imaged spectral windows contain 1920 channels with a width of 122~\mbox{kHz} (0.11 km s$^{-1}$). The observed phase centers are $\alpha_\mathrm{J2000}:$ 19:37:00.89, $\delta_\mathrm{J2000}:$ +07:34:09.5 for B335, and $\alpha_\mathrm{J2000}:$ 18:17:29.91, $\delta_\mathrm{J2000}:$ -04:39:39.6 for L483. The source velocities applied to the spectral tunings are listed in Table \ref{table:observations}.

Each dataset was pipeline-calibrated using {\sc casa 5.6} \citep{casa}. Subsequent phase self-calibration was attempted for each dataset with {\sc casa 5.6}. For B335, self-calibration was successful with an improved signal-to-noise ratio (S/N) of $\sim 25\%$; however, for L483, no improvement in S/N level was achieved, and therefore the ALMA pipeline product was preferred.

Imaging was performed with the {\sc tclean} algorithm with a robust parameter of 0.5. Toward each source, a continuum image at 302~GHz was created and compared with previous ALMA observations toward the sources. The intensities and morphologies of the continuum images are consistent with previous interferometric observations of the sources \citep[e.g.,][]{imai2016, oya2017, jacobsen2018, imai2019}. The final continuum images are shown in Fig. \ref{fig:cont}.
The synthesized beam size in spectral window centered on the D$_2$O transition are $0.\!\!^{\prime\prime}38\times0.\!\!^{\prime\prime}37    $ and $0.\!\!^{\prime\prime}56\times0.\!\!^{\prime\prime}46$ for B335 and L483, respectively.

Continuum subtraction for each spectral window was performed using the sigma-clipping method with the {\sc statcont} code presented in detail in \cite{statcont}. We preferred to use this method because both sources display rich hot corino emission, which made continuum subtraction through the {\sc uvcontsub} routine in {\sc casa} challenging \citep[see, e.g., ][]{jorgensen2016}. For the line analysis, we extracted the spectrum from a single pixel toward the continuum peak position for each source.

The emission lines presented in this work are identified using data from the Jet Propulsion Laboratory \citep[JPL,][]{JPL} and the Cologne Database for Molecular Spectroscopy \citep[CDMS,][]{CDMS}. The spectroscopic data used to identify D$_2$O, HDO, H$_{2}^{18}$O are from \cite{D2O_ref}, \cite{hdo_ref}, and \cite{h218o_ref}, respectively. Furthermore, spectroscopic data for CH$_3$OCH$_3$, and CH$_3$OD are from \cite{CH3OCH3_ref} and \cite{CH3OD_ref}, respectively. For CH$_3$OD, the exact partition function is unknown, and the current analysis relies on a partition function scaled from the CH$_{3}^{18}$OH partition function, similar to \cite{jorgensen2018}.
All querying was done through the Splatalogue interface\footnote{\url{https://splatalogue.online}}. Throughout this work, the H$_2$O column density is estimated from H$_{2}^{18}$O by adopting a ${}^{16}$O/${}^{18}$O ratio of 560 \citep{wilson1994}.

\section{Results} \label{sec:3}
The D$_2$O lines were detected toward both sources as part of a broad emission feature. The emission feature was identified as D$_2$O blended with two CH$_3$OD transitions and, potentially, one CH$_3$OCH$_3$ transition. This identification is consistent with the observed feature toward IRAS~2A \citep{coutens2014}. As the line is heavily blended, estimation of the column density is challenging, and two complimentary methods were employed. The spatial extent of the D$_2$O and CH$_3$OD emission is confined to the inner envelope, as shown in Fig. \ref{fig:moment0}. 

\begin{figure*}[!htb]
   \centering
   \includegraphics[width=\hsize]{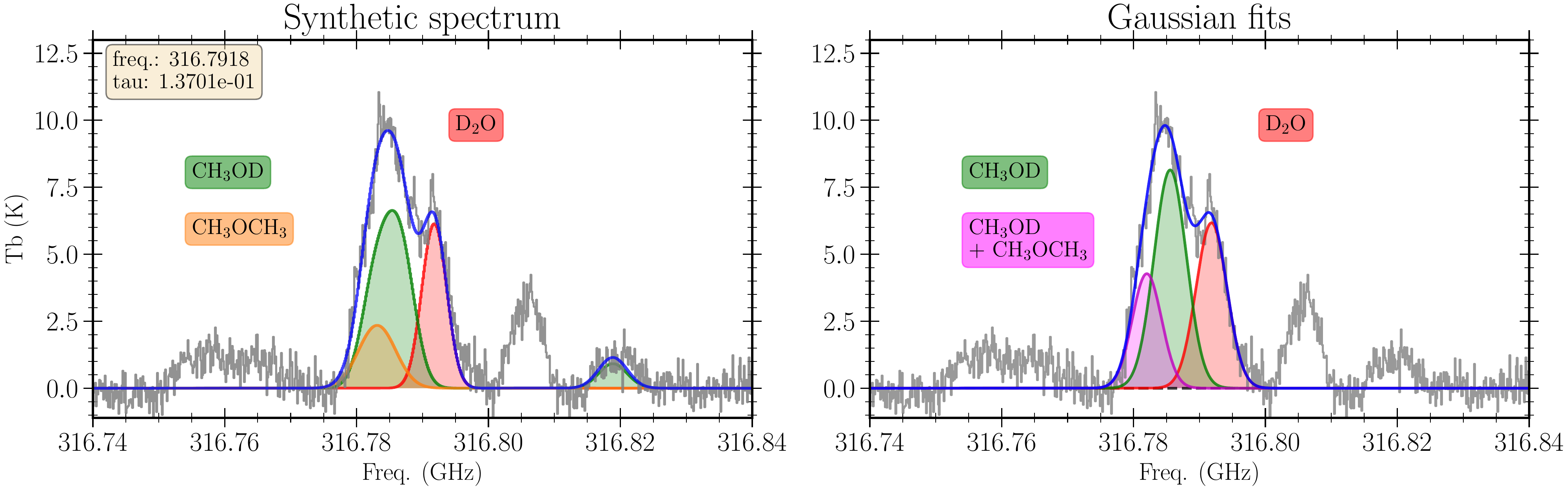}
      \caption{Overview of the two methods used to determine the column density of D$_2$O toward B335. \emph{Left}) Synthetic spectrum model for the emission feature toward B335 for an assumed source size of 0$.\!\!^{\prime\prime}$2. The estimated optical depth $\tau$ for D$_2$O is shown in the upper left corner. The synthetic spectrum includes components from D$_2$O, CH$_3$OD, and CH$_3$OCH$_3$. \emph{Right}) Least-squares fit to the emission profile. The best fit includes three Gaussian components, one identified as D$_2$O, one identified as CH$_3$OD at 316.7651~GHz, and a third component that may be either CH$_3$OD at 316.7916~GHz or CH$_3$OCH$_3$ at 316.7904~GHz, or a convolution of both.}
         \label{fig:B335_spectra}
\end{figure*}

\begin{figure*}
   \centering
   \includegraphics[width=\hsize]{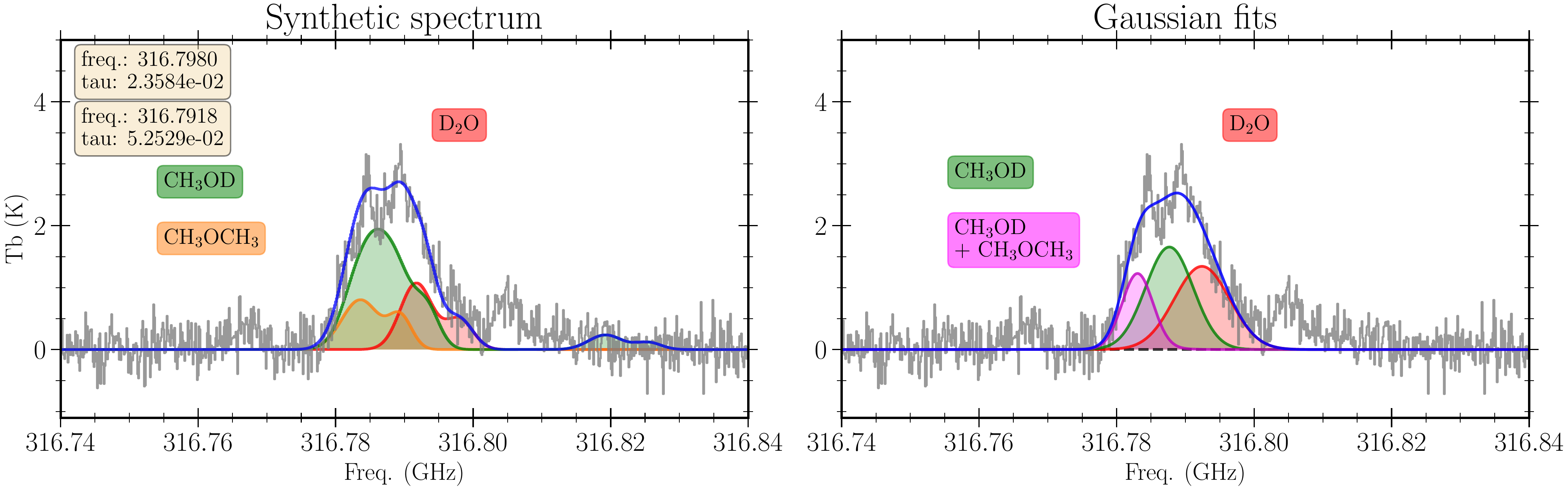}
      \caption{Overview of the two methods used to determine the column density of D$_2$O toward L483. \emph{Left}) Synthetic spectrum model for the emission feature toward L483 for an assumed source size of 0$.\!\!^{\prime\prime}$2. All molecules are fitted with double Gaussians. The estimated optical depth $\tau$ for both D$_2$O profiles are shown in the upper left corner. The synthetic spectrum includes components from D$_2$O, CH$_3$OD, and CH$_3$OCH$_3$. \emph{Right}) Least-squares fit to the emission profile. The best fit includes three components, one identified as D$_2$O, one identified as CH$_3$OD at 316.7651~GHz, and a third component that may be either CH$_3$OD at 316.7916~GHz or CH$_3$OCH$_3$ at 316.7904~GHz, or a convolution of both.}
         \label{fig:L483_spectra}
\end{figure*}

\subsection{Synthetic spectrum modeling}
The first method for the estimation of the D$_2$O column density relies on a synthetic spectrum of the sources. Combining the ALMA data from \cite{jensen2019} with those presented in this work, spectral coverage extends to several spectral windows in ALMA bands 5, 6, and 7. For B335, two unblended transitions of CH$_3$OD and four unblended transitions of CH$_3$OCH$_3$ were identified. For L483, three unblended transitions of CH$_3$OD and three unblended transitions of CH$_3$OCH$_3$ were identified. This allowed for accurate modeling of the blended emission feature around the D$_2$O line (i.e., the combined emission from D$_2$O, CH$_3$OD, and CH$_3$OCH$_3$). The best fit was determined by $\chi^2$ minimization of the data around emission lines. Blended lines and transitions with high optical thickness were manually excluded from the $\chi^2$ estimation. The synthetic spectrum was calculated following the methodology presented in \citet{moller2017} with the addition of an optical depth correction from \citet{goldsmith1999}.
Toward B335, the observed spectrum was best reproduced by a synthetic model with $N$(D$_2$O) = $3.5\times10^{15}$~cm$^{-2}$, $N$(CH$_3$OD) = $2.5\times10^{17}$~cm$^{-2}$, and $N$(CH$_3$OCH$_3$) = $1.3\times10^{17}$~cm$^{-2}$. 
For the water isotopologs, the synthetic spectrum model yields HDO/H$_2$O $= (6.3\pm1.5)\times10^{-3}$ and D$_2$O/HDO $= (1.4\pm0.1)\times10^{-2}$ for B335.

For L483, the synthetic spectrum modeling yields column densities of $N$(D$_2$O) = $1.1\times10^{15}$~cm$^{-2}$, $N$(CH$_3$OD) = $1.7\times10^{17}$~cm$^{-2}$, and $N$(CH$_3$OCH$_3$) = $7.9\times10^{16}$~cm$^{-2}$. The D/H ratios toward this source are then HDO/H$_2$O $= (4.0\pm0.5)\times10^{-3}$ and D$_2$O/HDO $= (1.0\pm0.2)\times10^{-2}$.
The synthetic spectra are presented in the left panels of Figs. \ref{fig:B335_spectra} and \ref{fig:L483_spectra}. For L483, the best-fit spectral model shows some deviation from the observed profile, however, the D$_2$O column density is robust: changing the parameters of the CH$_3$OD and CH$_3$OCH$_3$ components has little impact on the D$_2$O abundance. We note that for L483, many emission lines are double-peaked Gaussians \citep[see, e.g., ][]{oya2017, jacobsen2018}. The parameters for the best-fit synthetic spectrum are presented in Table \ref{table:fit}, for B335 and L483. 
The full spectral models for both sources are shown in Appendix \ref{app:fullspectrum}.

\subsection{Least-squares fitting}
In the second, complimentary method, the line profiles were fitted using the least-squares method with the {\sc scipy} routine {\sc curve\_fit}. Toward B335, the fit was allowed to include up to four Gaussian line profiles, one for each potential transition in the blended emission feature. Bounds on the fits were estimated from the HDO data, and other non-blended emission lines in the dataset. The best fit does not include a fourth component, corresponding to either the CH$_3$OCH$_3$ line at 316.7904~GHz or the CH$_3$OD transition at 316.7916~GHz (i.e., only three Gaussian profiles were included in the final fit). The exclusion of a fourth component has limited impact on the Gaussian profile for D$_2$O. The resulting fit is shown the right panel of Fig. \ref{fig:B335_spectra}. The column density is estimated from the fitted line profile by assuming local thermodynamic equilibrium (LTE) \citep{mangum2015}. We adopted an excitation temperature of 220~K, which is derived from the synthetic spectrum; this method results in a D$_2$O column density of $4.0\times10^{15}$~cm$^{-2}$. The HDO transitions presented in \cite{jensen2019} give a HDO column density of $1.9\times10^{17}$~cm$^{-2}$ at the same excitation temperature, when a beam filling factor is applied for a source size of $0.\!\!^{\prime\prime}2$. This method then results in a D$_2$O/HDO ratio of $(2.1\pm0.6)\times10^{-2}$. Evidently, the D$_2$O/HDO ratios toward B335 are somewhat higher when using the Gaussian fits, however both methods yield high D$_2$O/HDO ratios $\gtrsim 10^{-2}$. We note that the HDO column density is different from that presented in \cite{jensen2019} because a higher excitation temperature of 220~K and a smaller source size are used in this work. Furthermore, the synthetic model includes an optical depth correction and identifies a slight blending of the H$_{2}^{18}$O line, increasing the HDO/H$_2$O ratio. Because of this, we consider the column densities presented in this work as more accurate. However, this does not impact the results and discussions of the relative HDO/H$_2$O abundances in \cite{jensen2019}, in which identical excitation temperatures were used for all sources. The differences in the reported HDO/H$_2$O ratios in this work and in \citet{jensen2019} illustrate the dependence on the methodology and assumptions, however, the observed dichotomy between isolated and clustered protostars reported in that paper is robust. To remove the observed dichotomy requires consistently adopting lower excitation temperatures for the isolated sources, independent of the luminosity of the respective sources.

In the case of L483, the emission lines are generally double-peaked Gaussian profiles \citep{jensen2019}. These features are more challenging to fit when strongly blended and attempts to fit double-peaked Gaussians to all four potential emission lines did not produce good results. Instead, single Gaussians were fitted to each transition, which provided reasonable fits (as shown in the right panel of Fig. \ref{fig:L483_spectra}). The fitted Gaussian profiles result in $N$(D$_2$O) = $1.7\times10^{15}$~cm$^{-2}$, which, combined with $N$(HDO) = $1.2\times10^{17}$~cm$^{-2}$, yield D$_2$O/HDO $= (1.4\pm0.4)\times10^{-2}$. Again, this method returns a higher D$_2$O/HDO than the synthetic spectrum model. For L483, an excitation temperature of 165~K was used, which was derived from the synthetic spectrum model.
Overall, both methods agree, given the inherent flux density uncertainty of ALMA ($\sim10\%$) and the uncertainties in the fitting ($\sim10\%$). 

Both methods assume that water emission from the warm envelope is in LTE. Previous studies of water isotopologs with interferometers have assessed the validity of this approach. \cite{persson2014} compare the optically thin LTE approximation to radiative transfer models for two Class 0 sources and find agreement between the two methods. \cite{coutens2014} compare non-LTE {\sc radex} calculations and LTE calculations for the same HDO transitions presented in this work and found that the LTE approach is valid for these transitions. Generally, the LTE approach is suitable for water emission in the inner envelope, where the densities are similar or higher than the critical densities for the targeted transitions.

\begin{table*} \label{table:fit}
\centering\caption{Overview of the column densities for D$_2$O, HDO, and H$_{2}^{18}$O derived with the best-fit synthetic spectrum model ($N_\mathrm{model}$) and with the fitted Gaussian profiles ($N_\mathrm{fit}$). The reported full width at half maximum (FWHM) and $v_\mathrm{LSR}$ are for the synthetic spectrum model. A source size of 0$.\!\!^{\prime\prime}$2 is assumed for both sources for all molecules.}    
\begin{tabular}{c c c c c c}
\hline \hline    
\noalign{\smallskip} 
Molecule & $N_\mathrm{model}$ (cm$^{-2}$) & $N_\mathrm{fit}$ (cm$^{-2}$) & FWHM (km/s) & $v_\mathrm{LSR}$ (km/s) & $T_\mathrm{ex}$ (K) \\
\noalign{\smallskip} 
\hline
\noalign{\smallskip} 
\multicolumn{6}{c}{\it L483} \\
\hline
\noalign{\smallskip} 
  D$_2$O & $1.1\times10^{15}$   & $1.7\times10^{15}$   &  (4.5, 5.2)  &   (1.75, 7.65)   &  165   \\
  HDO &   $1.1\times10^{17}$   & $1.2\times10^{17}$   &  (4.9, 5.7)  &    (1.75, 7.65)   &  165   \\
  H$_{2}^{18}$O &   $5.1\times10^{16}$   & $5.9\times10^{16}$   &  (6.2, 7.1)  &     (1.75, 7.65)   &  165   \\
  \noalign{\smallskip} 
  \hline
  \noalign{\smallskip} 
\multicolumn{6}{c}{\it B335} \\
\hline
\noalign{\smallskip} 
  D$_2$O & $3.5\times10^{15}$   &$4.0\times10^{15}$   &  4.25  &    7.6   &  220   \\
  HDO &   $2.5\times10^{17}$   &  $1.9\times10^{17}$   & 4.0  &    7.9   &  220   \\
  H$_{2}^{18}$O &   $6.3\times10^{16}$   & $7.1\times10^{16}$   &  5.65  &    7.9   &  220   \\
  \noalign{\smallskip} 
\hline
\end{tabular}
\end{table*}

 

\section{Discussion}  \label{sec:4}
The observed sources show high abundances of doubly deuterated water, D$_2$O, with D$_2$O/HDO $\gtrsim$ $10^{-2}$ and D$_2$O/HDO > HDO/H$_2$O for both sources. The derived f$_\mathrm{D2}$/f$_\mathrm{D1}$ ratios for L483 and B335 are $\sim$2.5 and $\sim$2, respectively. This is similar to the previous determination of D$_2$O/HDO in the warm envelope toward IRAS~2A presented in \cite{coutens2014}, who reported D$_2$O/HDO $\gtrsim$ $10^{-2}$ and f$_\mathrm{D2}$/f$_\mathrm{D1} \gtrsim 7$. Table \ref{table:deuteration} presents an overview of the HDO/H$_2$O and D$_2$O/HDO ratios toward low-mass Class 0 sources, limited to data which are determined with interferometric observations of the warm envelope. 
The high D$_2$O abundances reported in this work has several implications for the inheritance of water and the tentative correlation between protostellar environment and D/H ratio in water, which is discussed below.

\begin{table*}
\caption{Overview of the D/H ratios of water in the warm region ($\lesssim 100~$au, $T \gtrsim 100$~K) toward Class 0 protostars. For L483 and B335, the reported values are derived from the synthetic spectrum model.}             
\label{table:deuteration}      
\centering 
\smallskip \smallskip

\begin{tabular}{c c c c c}        
\hline\hline                 
            \noalign{\smallskip} 

Object & HDO/H$_2$O ($\times10^{-4}$) &  D$_2$O/HDO ($\times10^{-2}$) & f$_\mathrm{D2}$/f$_\mathrm{D1}$ & Reference \\
\hline                        
            \noalign{\smallskip} 
  \multicolumn{5}{c}{\it Clustered protostars} \\
  
\hline
NGC1333 IRAS 2A & $7.4\pm2.1$ & ... & ... & 1 \\
 -- & $17\pm8$ & $1.2\pm0.5$ & $7\pm4$ & 2 \\
NGC1333 IRAS 4A-NW & $5.4\pm1.5$ & ... & ... & 1,3 \\
NGC1333 IRAS 4B & $5.9\pm2.6$ &  ... & ... & 1 \\
IRAS 16293--2422 & $9.2\pm2.6$ &   ... & ... & 1 \\
\hline                        
            \noalign{\smallskip} 
  \multicolumn{5}{c}{\it Isolated protostars} \\
\hline
BHR71--IRS1 & $18\pm4$ & ... & ... & 3 \\
B335 & $17\pm3$ & ... & ...  & 3 \\
-- & $63\pm15$ & $1.4\pm0.1$ & $2.2\pm0.5$  & 4 \\
L483 & $22\pm4$ & ... & ... & 3 \\
-- & $40\pm5$ & $1.0\pm0.2$ & $2.5\pm0.3$ & 4 \\
	\noalign{\smallskip} 
	\hline                        
\end{tabular}
\tablefoot{The table only include interferometric observations of the inner warm region toward low-mass Class 0 protostars.}
\tablebib{
(1) \citet{persson2014}; (2) \citet{coutens2014}; (3) \citet{jensen2019}; (4) This work.
   }
\end{table*}

\subsection{Inheritance of water D/H ratio}
The D/H ratio of water is an essential tool for determining the chemical evolution of water during star and planet formation. Deuterium fractionation of water can proceed through multiple pathways during the evolutionary stages of star and planet formation with varying efficiency \citep[e.g.,][]{caselli2012, dishoeck2014}. 
In a recent work, \cite{furuya2017} studied the evolution of the water D/H ratio during the star formation process from cloud to disk for hundreds of tracer particles in a 2D axisymmetric model, including both the protostellar collapse and the protoplanetary disk phases. These authors showed that water formation with HDO/H$_2$O ratios in the range $10^{-4}$--$10^{-3}$ can occur both through prestellar inheritance and in situ water formation in the cold midplane of the protoplanetary disk. Hence, distinguishing between inheritance and in situ formation based solely on the HDO/H$_2$O ratio is challenging. 
Meanwhile, models suggest that the formation of D$_2$O is inefficient in protoplanetary disks when compared to the prestellar core phase. Therefore, \cite{furuya2017} suggest that the ratio of f$_\mathrm{D2}$/f$_\mathrm{D1}$ is a more robust tracer of chemical inheritance for water. Quantitatively, the model of \cite{furuya2017} shows that a f$_\mathrm{D2}$/f$_\mathrm{D1}$ ratio around $\sim10$ is formed in the prestellar phase, while a much lower ratio of f$_\mathrm{D2}$/f$_\mathrm{D1} \sim0.1$ is found for water formed within the protoplanetary disk. 

For L483 and B335, the observed f$_\mathrm{D2}$/f$_\mathrm{D1}$ ratios are $\gtrsim$2. Prior to this work, f$_\mathrm{D2}$/f$_\mathrm{D1}$ was determined in the hot corino toward IRAS~2A, where the ratio is $\gtrsim7$. The observed ratios, f$_\mathrm{D2}$/f$_\mathrm{D1} > 1$, suggest that water in the warm envelope is predominantly inherited from the prestellar phase and has not undergone reprocessing during accretion from the cloud to the warm envelope toward these sources. 

Recently, \cite{jensen2021} studied the D/H ratio of water during the formation of several low-mass protostars in a dynamic 3D magnetohydrodynamics simulation of a molecular cloud region. For each protostar, thousands of tracer particles recorded the physical evolution during the accretion process from the large-scale environment down to the warm envelope. The temperature structure of the envelopes was calculated using the Monte Carlo radiative transfer code {\sc radmc-3d} \citep{dullemond2012}. Subsequently, the chemical evolution along each particle trajectory was modeled to determine the gas-phase f$_\mathrm{D2}$ and f$_\mathrm{D1}$ ratios in the warm envelope. These authors find that f$_\mathrm{D2}$/f$_\mathrm{D1}$ depends on the choice of initial conditions in the prestellar cloud (e.g., the duration and temperature of this phase). A similar conclusion is presented in \cite{taquet2013model}, in which the authors study the formation of deuterated water for a broad range of initial conditions in the prestellar phase. These results suggest that the lower f$_\mathrm{D2}$/f$_\mathrm{D1}$ ratios may stem from different prestellar conditions, which was not included in the study by \cite{furuya2017}. 
By studying a range of initial conditions, \cite{jensen2021} show that the D/H ratio of water in the warm envelope is determined by the conditions in the prestellar phase and is not sensitive to variations during the protostellar collapse itself, in other words, suggestive of chemical inheritance of water from the prestellar stages. A high degree of inheritance for water is also suggested by \cite{cleeves2014}, who show that the D/H ratio of water synthesized in situ in protoplanetary disks is low, and find that a high  degree of inheritance is necessary to explain the D/H ratios of comets in the Solar System with low ionization rates. 


With the high observed D$_2$O/HDO ratios in the warm envelope, the case for a high degree of inheritance from the prestellar core stage down to the warm envelope ($\lesssim 100~$au) appears strong. 
The physical and chemical link between the warm envelope and the formation of protoplanetary disks and planets is unclear \citep[e.g.,][]{harsono2020}.
Recently, the evidence for early planet formation is growing with a number of detections of planetary signatures in young systems \citep[e.g.,][]{keppler2018, teague2018, harsono2018, pinte2020, alves2020}. Furthermore, advances in planet formation theory suggest that planet formation may be a fast process and can occur early \citep[e.g.,][]{johansen2017}. If planet formation starts early, the chemical inheritance observed in these Class 0 sources may directly impact the chemistry at the onset of planet formation. Currently, the D/H ratio of water in bona fide protoplanetary disks is unknown.
An alternative tracer for the chemistry during planet formation is the study of the chemical composition of comets, including the D/H ratio \citep[e.g.,][]{mumma2011, bockelee2015}. Comets are planetesimals in the outer solar nebula that record the pristine ice composition in the Solar System during planet formation.
Tentative evidence for a chemical link between the warm envelope and protoplanetary disks is emerging. \cite{altwegg2017} determine the D/H ratios of water in the coma of comet 67P/Churyumov-Gerasimenko (hereafter comet 67P) with the ROSINA instrument. They found  D$_2$O/HDO~$= (1.8\pm0.9)\times10^{-2}$ and HDO/H$_2$O~$= (1.05\pm0.14)\times10^{-3}$, corresponding to f$_\mathrm{D2}$/f$_\mathrm{D1} \sim 17$, which is incompatible with in situ D$_2$O formation in the protoplanetary disk \citep{furuya2017}. This suggests a high degree of inheritance, and furthermore the measured D$_2$O/HDO ratio is similar to the ratios observed in warm envelopes. Overall, current estimates of the D/H ratios of water in comets and warm envelopes could indicate a similar chemical evolution for the cold icy component that constitute comets such as comet 67P, and the material that enters the warm envelope, where ice is sublimated off the dust grains. A chemical link between the warm envelope and cometary bodies is also supported by the apparent correlation between complex organic molecules (COMs) in the hot corino of IRAS16293--2422 and comet 67P, which may indicate a link between the chemical inventory in cold icy planetesimals and hot corinos \citep{drozdovskaya2019}. Similarly, a recent study by \cite{bianchi2019} also finds a correlation between the abundances of COMs in protostars and comets. The authors compared the inventory of COMs for a larger sample of Class 0 and Class I sources to cometary values in the literature and find good agreement between the abundances of COMs in Class 0 sources and comets, while the Class I sources showed more variation. 

\subsection{Environmental impact on the D/H ratio}
The D$_2$O/HDO ratios toward B335 and L483 are on the order of $10^{-2}$, similar to the lower limit derived in \cite{coutens2014} toward IRAS~2A. Following the definitions of \cite{jensen2019}, B335 and L483 are isolated protostars (i.e., not associated with any cloud complexes) while IRAS~2A is located in NGC1333. Given the higher HDO/H$_2$O ratios in the isolated sources, these results appear to follow the same trend between water deuterium fractionation and protostellar environment. The isolated protostars show higher D/H ratios in HDO and similar D$_2$O/HDO ratios, that is, the overall D/H ratio is higher than in clustered counterparts. Another way to illustrate this is to look at the D$_2$O/H$_2$O ratio. For L483, B335, and IRAS~2A, the D$_2$O/H$_2$O is $9\times10^{-5}$, $4\times10^{-5}$, and $2\times10^{-5}$, respectively. Hence, the D$_2$O/H$_2$O ratios toward the isolated protostars are a factor $\gtrsim2$ higher than the clustered source observed so far. The D$_2$O abundances therefore corroborate a higher D/H ratio of water toward isolated protostars, similar to the trend observed in HDO. However, the sample size remains too small to draw definite conclusions and should be further expanded. 

\cite{jensen2019} propose that the observed trend in D/H ratios could originate from either higher temperatures or stronger irradiation in star-forming clusters, or a difference in the collapse timescale between isolated cores, which is subject to limited external perturbations, and clustered sources for which several mechanisms may perturb the dynamics of the cores, such as jets and outflows from nearby stars. Clustered star-forming regions may also be intrinsically more turbulent; that is, the probability of multiple star formation events may correlate with the turbulence in the region \citep[e.g.,][]{ward-thompson2007, krumholz2014}.

The physicochemical 3D models of water deuteration presented in \cite{jensen2021} find that variations in the temperature or duration of the prestellar phase can induce the observed dichotomy in water D/H ratios. Higher temperatures during both the prestellar core phase and protostellar collapse lead to lower D/H ratios of water. Similarly, a shorter embedded phase (i.e., a shorter prestellar core duration) leads to a lower D/H ratio, which is consistent with the hypothesis that clustered sources collapse on shorter timescales. Both effects can drive the differences in D/H ratios between clustered and isolated low-mass protostars. Meanwhile, variations in irradiation appear unlikely to directly drive the dichotomy in observed D/H water ratios in the warm envelope toward low-mass protostars. The study by \cite{jensen2021} demonstrates a link between the chemistry in the  prestellar stages and gas-phase water D/H ratio in the warm envelope, which is corroborated by the observed D/H ratios of water from this region toward Class 0 sources. This link motivates further study, both observationally and theoretically, to determine whether other molecules exhibit a similar link between conditions in the environment and the chemistry in the warm envelope region toward young embedded protostars.

\section{Summary and outlook}  \label{sec:5}
This work presents the first ALMA observations of D$_2$O in the  warm envelope toward low-mass protostars. We determine HDO/H$_2$O and D$_2$O/HDO ratios by combining the column densities of D$_2$O with HDO and H$_{2}^{18}$O data. These ratios provide new insights into the chemical evolution of water during the star formation process.
\begin{itemize}
    \item Both L483 and B335 have D$_2$O/HDO ratios $\gtrsim 10^{-2}$. When combined with previous ALMA estimates of the HDO/H$_2$O ratios, the [D$_2$O/HDO]/[HDO/H$_2$O] ratios are derived to be $\sim$2.5 and $\sim$2, respectively. D/H ratios of this magnitude are strong evidence of chemical inheritance from the prestellar core stage. Chemical models suggest that a D$_2$O/HDO ratio of this magnitude can only form during the cold phase prior to star formation. 
    \item Both sources are isolated and show enhanced D$_2$O/H$_2$O ratios compared to a clustered counterpart by a factor of $\gtrsim 2$. The high D$_2$O/H$_2$O ratios strengthen the evidence for a correlation between the D/H ratio of water and the protostellar environment, as suggested in \cite{jensen2019}. This is consistent with a high degree of inheritance from the prestellar phases and provides a mechanism for chemical diversity in the envelopes surrounding embedded protostars (e.g., through a longer duration or colder temperatures in the prestellar phase).
    \item The observed D$_2$O/HDO ratios in hot corinos are similar to the value measured in Comet 67P of $\sim 10^{-2}$, providing tentative evidence of a common chemical evolution of the water that sublimates in the warm envelope and water incorporated in comets and icy planetesimals that are the building blocks of planets in the Solar System. 

\end{itemize}
These observations highlight the power of the deuterium fractionation as a tracer for the chemical trail during star formation. More observations of the D/H ratios of water at different stages of star and planet formation will provide further constraints on the amount of inheritance between the evolutionary stages and establish the impact of the local environment on the chemical evolution. The relation between the local cloud environment and isotope fractionation should be explored further by studying, for example, nitrogen or carbon fractionation, toward both isolated and clustered sources at various evolutionary stages.

\begin{acknowledgements}
The authors thank the referee for useful comments that helped improve the manuscript.
This paper makes use of the following ALMA data: ADS/JAO.ALMA\#2017.1.00693.S, ADS/JAO.ALMA\#2019.1.00720.S.
ALMA is a partnership of ESO (representing its member states), NSF (USA) and NINS (Japan), together with NRC (Canada), NSC and ASIAA (Taiwan), and KASI (Republic of Korea), in cooperation with the Republic of Chile. The Joint ALMA Observatory is operated by ESO, AUI/NRAO and NAOJ.
	The group of JKJ acknowledges support from the European Research Council (ERC) under the European Union's Horizon 2020 research and innovation programme (grant agreement No 646908) through ERC Consolidator Grant "S4F". AC acknowledges financial support from the Agence Nationale de la Recherche (grant ANR-19-ERC7-0001-01). This paper makes use of {\sc matplotlib} \citep{hunter2007} and {\sc scipy} \citep{2020SciPy-NMeth}.
\end{acknowledgements}

%
%

\bibliographystyle{aa}
\bibliography{D2O.bib}
\begin{appendix} 
\section{Full synthetic models}\label{app:fullspectrum}
\begin{figure*}
   \centering
   \includegraphics[width=\hsize]{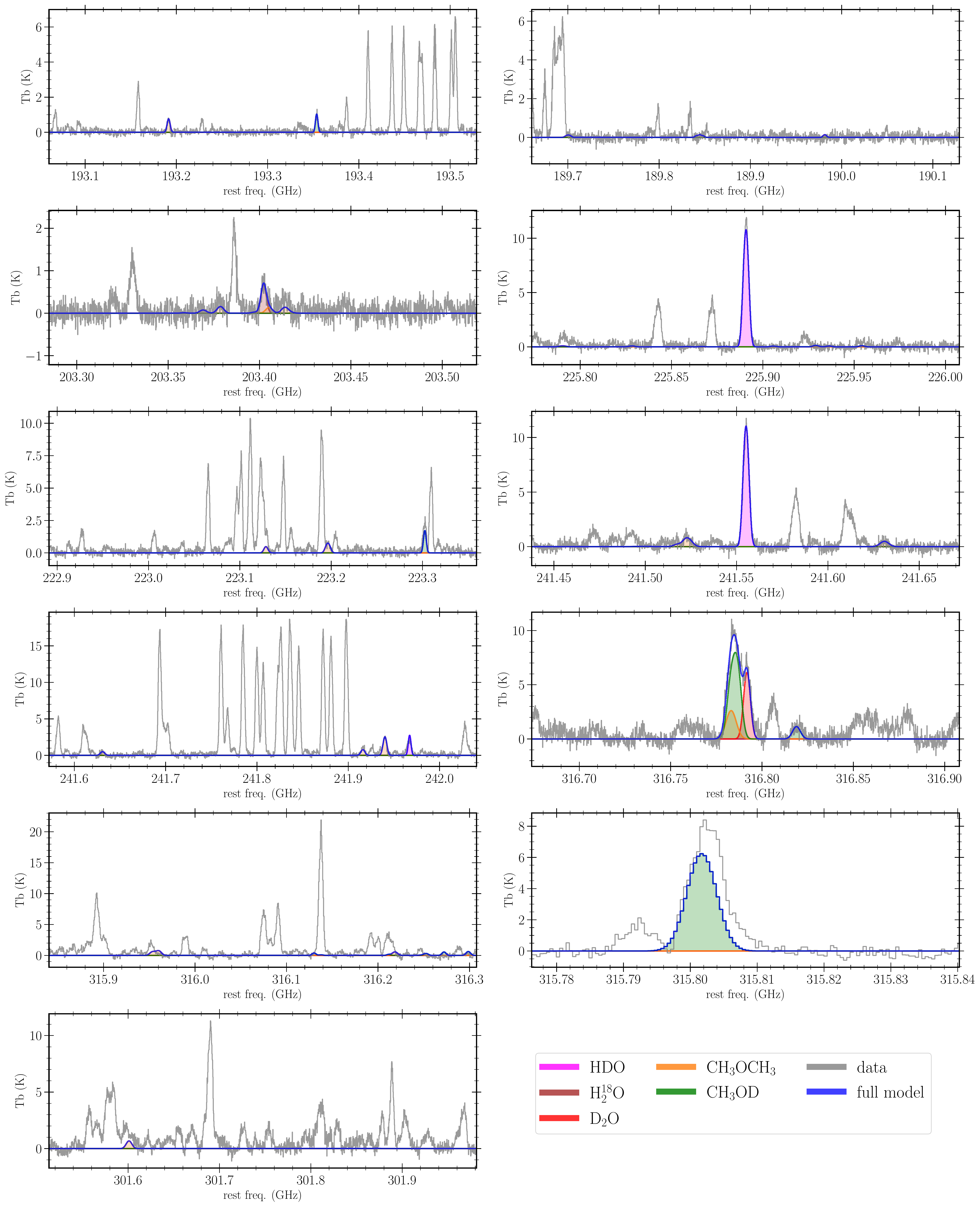}
      \caption{Overview of the complete synthetic spectrum model for emission lines toward B335 for an assumed source size of 0$.\!\!^{\prime\prime}$2. The full model includes H$_{2}^{18}$O, HDO , D$_2$O, CH$_3$OD, and CH$_3$OCH$_3$.}
         \label{fig:B335_spectrum1}
\end{figure*}

\begin{figure*}
   \centering
   \includegraphics[width=\hsize]{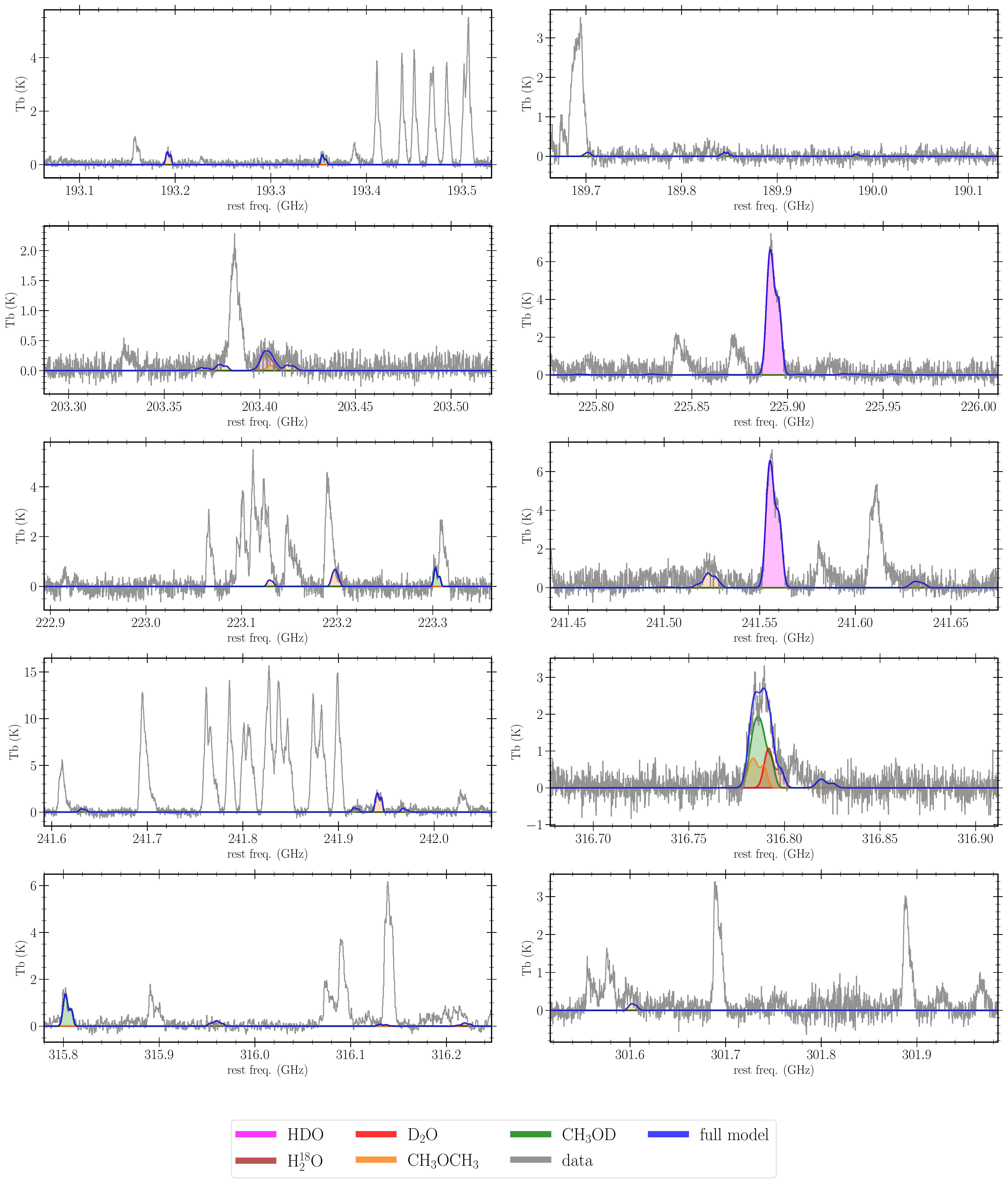}
      \caption{Overview of the complete synthetic spectrum model for emission lines toward L483 for an assumed source size of 0$.\!\!^{\prime\prime}$2. The full model includes H$_{2}^{18}$O, HDO , D$_2$O, CH$_3$OD, and CH$_3$OCH$_3$.}
         \label{fig:L483_spectrum1}
\end{figure*}

The complete synthetic spectrum models for H$_{2}^{18}$O, HDO , D$_2$O, CH$_3$OD, CH$_3$OCH$_3$ are shown in the Figs. \ref{fig:B335_spectrum1} and \ref{fig:L483_spectrum1}. 


\end{appendix}
\end{document}